\documentclass[aps,prl,showpacs]{revtex4}
\usepackage{graphicx}
\usepackage{verbatim}

\newcommand{\req}[1]{(\ref{#1})}
\newcommand{\be}{\begin{equation}}
\newcommand{\ee}{\end{equation}}
\newcommand{\bea}{\begin{eqnarray}}
\newcommand{\eea}{\end{eqnarray}}
\newcommand{\dd}{\textrm{d}}
\newcommand{\pr}[1]{\left(#1\right)}
\newcommand{\cro}[1]{\left[#1\right]}

\newcommand{\avg}[1]{\langle{#1}\rangle}
\newcommand{\ovl}[1]{\overline{#1}}
\newcommand{\BE}{\begin{eqnarray}}
\newcommand{\EE}{\end{eqnarray}}
\newcommand{\BEn}{\begin{eqnarray*}}
\newcommand{\EEn}{\end{eqnarray*}}
\newcommand{\barr}{\begin{array}}
\newcommand{\earr}{\end{array}}

\newcommand{\bit}{\begin{itemize}}      
\newcommand{\eit}{\end{itemize}}
\newcommand{\bc}{\begin{center}}
\newcommand{\ec}{\end{center}}
\newcommand{\ben}{\begin{enumerate}}    
\newcommand{\een}{\end{enumerate}}

\newcommand{\eps}{\epsilon}

\begin{document}

\title{Minority Games with heterogeneous timescales}
\author{Giancarlo Mosetti$^1$, Damien Challet$^2$, Yi-Cheng Zhang$^1$}
\affiliation{$^1$D\'epartement de Physique, Universit\'e de Fribourg, Chemin du Mus\'ee 3, 1700 Fribourg, Switzerland \\
$^2$Nomura Centre for Quantitative Finance, Mathematical Institute, Oxford University, 24--29 St Giles', Oxford OX1 3LB, United Kingdom}
\email{challet@maths.ox.ac.uk}
\date{\today}


\begin{abstract}
Minority games where groups of agents remember, react or incorporate information with different timescales are investigated. We support our findings by analytical arguments whenever possible.
\end{abstract}

\pacs{88}
\maketitle

\section{Introduction}

Heterogeneity is gradually being recognized as one of the most important ingredients for the modeling of financial markets. Among the many types of heterogeneities, timescales are difficult to understand analytically, because they increase usually much the complexity of the equations to solve. However, as argued very early \cite{DacoHeterogeneity}, financial market participants have very discernably different timescales, which deserve therefore a detailled study. Among recent works on the topic, two models of  stochastic volatility have explicitely included an infinite number of timescales \cite{BorlandBouchaud,ZumbachLynch}.

Here we investigate several types of timescale heterogeneities in the Minority Game model (MG) framework: strategy change frequency, strategy-strategy correlation, reaction rates and score memory. We are able to provide analytical support for our finding in the last three cases. Previous work addressed the effects of trading frequency \cite{Piai,ADMfrequency} analytically.
\section{Canonical MG}

 The minority game is easily defined: at each time step, all the players have to choose between two alternatives; those who happen to be in the minority win. This is variant of Arthur's El Farol bar \cite{Arthur} problem where the resource level is set to satisfy half of the people. The Minority Game is exactly solvable which makes it an ideal model to study and understand various aspects of the dynamics of competition. 

Mathematically, agent $i=1,\cdots,N$ takes action $a_i(t)\in\{-1,+1\}$ at each time step and receives a payoff $-a_i(t)A(t)$ where $A(t)$ is the aggregated outcome $A(t)=\sum_{i=1}^Na_i(t)$. The agents base their decisions on public information which is encoded in an integer number $\mu(t)$ 
drawn uniformly from ${1,\cdots,P}$. In order to process this information they are endowed with $S$ strategies, which are fixed maps, or look-up tables, from the current public information $\mu(t)$ to an action. At time $t$, agent $i$ decides to trust his best strategy $s_i(t)=\arg \max_{s=1,\cdots,S}y_{i,s}(t)$, where $y_{i,s}(t)$ is the score of strategy $s$ of agent $i$, which evolves according to
\be\label{scoreMG}
  y_{i,s}(t+1)=y_{i,s}(t)-a_{i,s}^{\mu(t)} A(t)
\ee

\subsection{Strategy change frequency}

The above definition of the MG assumes implicitely that all the agents may change the strategy that they use ($s_i(t)$) at each time step. Variants of the game where all the agents update synchronously their $s_i(t)$ every $T$ time steps have been studied in the literature \cite{Moro1,CoolenBatch,TobiasSherr,CDMP04}, and are exactly solvable in the limit $T\to\infty$, in which case they are called batch games. This part of the paper separates the populations into two groups. Fast agents behave as usual, whereas slow agents update their strategies synchronously every $T$ time steps. This introduces a heterogeneity of time scales. I n essence, it is similar to giving a longer history memory in games where the state $\mu(t)$ is the binary encoding of the last $M$ winning choice, also known as games with real histories. Giving a larger $M$ to a set of agents is already found in the very first paper on the MG, and was analyzed further in \cite{JohnsonEnhancedWinnings,MMM}. Such agents do surprisingly worse than their colleagues as long as the system is not deep into the symmetric phase, which is characterized by alternating winning sides; Metzler \cite{MetzlerAnti} showed that agents with a larger $M$ need a large alternating probability in order to be able to profit from it. The case we study here is much simpler as it does not require real histories and is {\em a priori} more suitable to mathematical understanding. Unfortunately, since the exact analytical solution does not exist for games where all the agents update synchronously their strategies every $T$ time steps, we cannot solve the mixed case either and must resort to numerical simulations.

The relative composition of the population is tuned by a parameter $\phi$: if $N$ is the total number of agents, $\phi N$ of them are ordinary Minority Game agents, i.e., fast, whereas $(1-\phi)N$ update their strategy choice variable $s_i(t)$ every $T$ time steps, that is, their scores evolve following
\bea\label{scoreMGT}
            y_i^s(t+1)&=&y_i^s(t)-\sum_{t'=t-T+1}^{t}a_{i,s}^{\mu(t')} A(t')  ~~{\rm if }~t~{\rm MOD}~T=0\nonumber\\     y_i^s(t+1)&=&y_i^s(t) ~~{\rm otherwise.}
\eea

As usual, we shall focus on the predictability 
\be
H = \frac{1}{P}\sum_{\nu=1}^P\avg{A|\nu}^2 
\ee
where $\avg{A|\nu}$ is the temporal average of $A$ conditional to $\mu(t)=\nu$. If $H>0$, knowing $\mu$ makes it possible to predict statistically the next outcome. $H$ measures the amount of information left by the agents in the game. The fluctuations $\sigma^2=\avg{A^2}$ play a special role as they quantify the quality of resource sharing achieved by the population, which is usually benchmarked against the fluctuations produced by random choice $\sigma^2/N=1$. It is easy to see indeed that the $\sigma^2$ is nothing else than the average total loss per time step of the population. Of particular interest in the case of competing populations are their respective average gains per time step.

\begin{figure}
\centerline{\includegraphics*[width=0.5\textwidth]{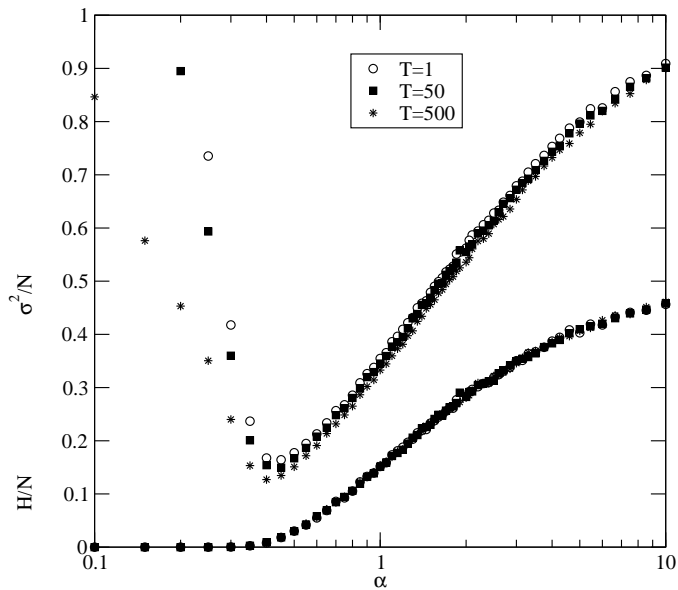}}
\caption{Variance $\sigma^2/N$ (top) and predictability $H/N$ (bottom) as a function of $\alpha$ for $\phi$=0.7 and several values of $T$. Simulations with $P=64$, averages over 200 samples}
\label{fig1}
\end{figure}

Figure \ref{fig1} reports the behaviour of $H$ and $\sigma^2$ as a function of $\alpha=P/N$, for a fixed 
$\phi$ and different values of $T$. In the predictable, asymmetric phase ($H>0$), the slow agents do not change the unique stationary state; as a result, $H$ does not depend on $T$, but $\sigma^2$ is slightly lowered as $T$ increases, because strategy switching occurs less often. On the other hand, in the unpredictable, symmetric phase ($H=0$), multiple stationary states exist as a result of broken ergodicity \cite{CMZe00,CoolenBatch}, and any modification to the system will accordingly change the final values of $\sigma$. This is the case here: the introduction of a small amount of slow  agents reduces the total amount of fluctuations because they damp the overreaction of the fast players without contributing too much to the global outcome $A$. 
\begin{figure}[htb]
\centerline{\includegraphics*[width=0.5\textwidth]{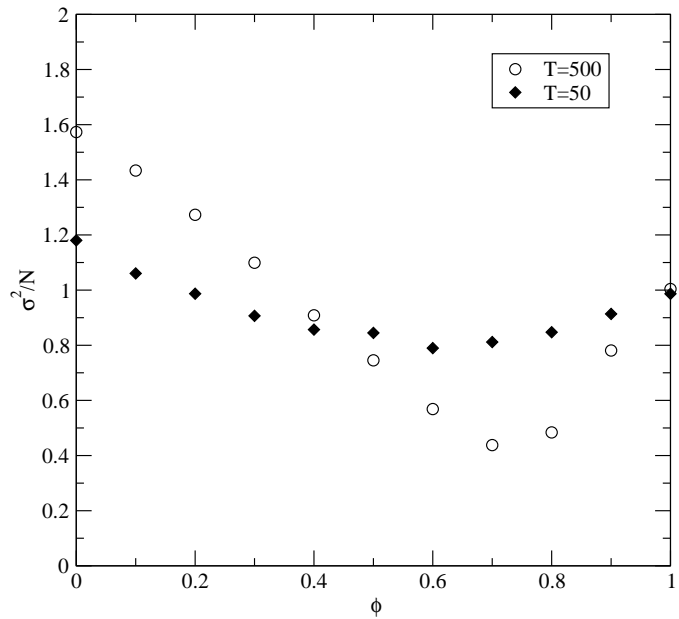}}
\caption{Volatility $\sigma^2/N$ as a function of  $\phi$ for $\alpha=0.2$ fixed ($P=32$, $N=160$).}
\label{fig2}
\end{figure}

What is more surprising is the behaviour of $\sigma^2$ in the symmetric phase when one varies $\phi$, as shown in Fig. \ref{fig2}: at fixed $\alpha=0.2$, slow agents ($\phi=0$) produce larger fluctuations than fast ones ($\phi=1$);  increasing the fraction of fast agents $\phi$ decreases the fluctuations which reaches a minimum below $1$, the random choice benchmark, and then increase again to reach the standard MG value, slightly above $1$. This means that the two groups live in symbiosis, and that there is a non-trivial optimal composition of the population. Other known examples of symbiosis in MGs include speculators and producers \cite{MMM}.

In such cases, it is natural to characterize the information ecology of the model (see \cite{ZMEM,MMM}), that is, who exploits who. To this end, Fig. \req{fig:article3} reports the average gain per time step of fast agents $\gamma_f$, which decreases monotonically as their concentration $\phi$ increases, but stays roughly constant as long as $\phi<0.4$. In this region, the slow agents provide information that the fast agents exploit. The losses of the latter are greatly reduced compared to $\phi=1$, but they seem not to be able to achieve positive gain on average. Similarly, when the slow agents are few ($\phi>0.8$), they profit from overreacting fast agents.


\begin{figure}[h!]
\centerline{\includegraphics*[width=0.8\textwidth]{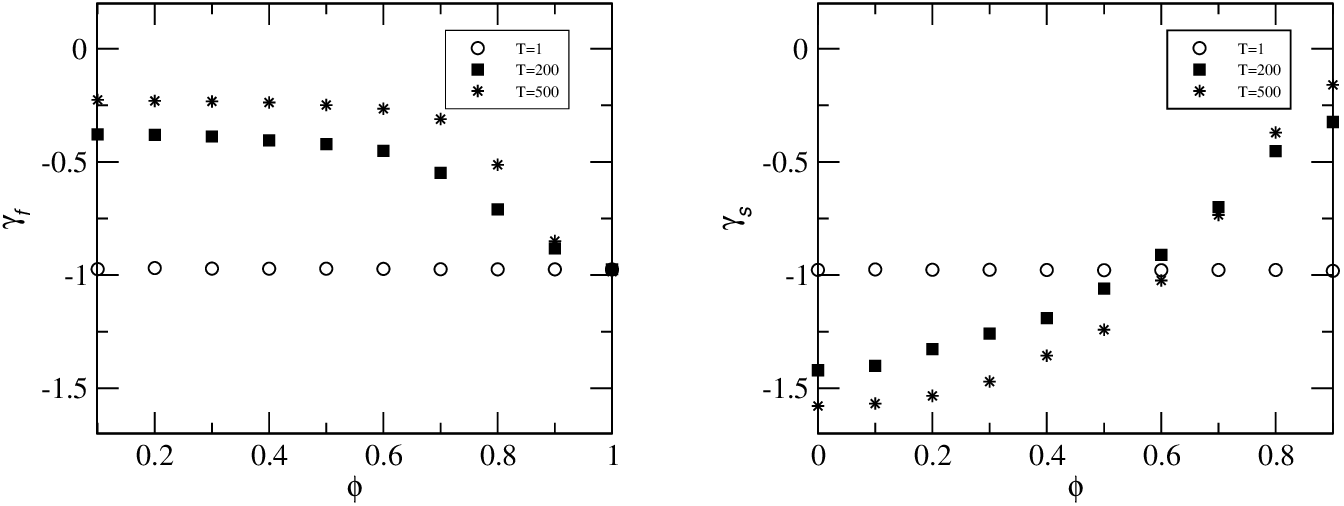}}
\caption{Average gain of fast (left) and slow (right) agents  as a function of $\phi$ for several $T$. Simulations with $N=300, P=64$, averages over 200 samples}
\label{fig:article3}
\end{figure}

\subsection{Strategy correlation} 

In the standard MG, all the $a_{i,s}^\mu$ are random variables, drawn completely independently from each other. Take an agent with two strategies; they will stipulate the same action a fraction $c=1/2$  of the  $\mu$s on average. In other words, the standard-MG agents behave in the same way irrespectively on their strategy choice for half of the market states $\mu$.\footnote{This is the origin of predictability in the MG \cite{CM99}.} MG with tunable $c$ where introduced in Ref. \cite{MMM} and also studied later in Refs \cite{Moro2,TobiasSherr}. The parameter $c$ induces a reaction time scale $\sim 1/(1-c)$: the smaller $c$,  the more adaptive an agent is; on the other end, people with $c=1$ are not adaptive: they inject predictability and are ideal candidates for exploitation. The latter were introduced as producers, that is, people who do not care much about timing in market, but use the market as a tool for exchanging goods \cite{ZMEM}; producers ($c=1$) and speculators ($c<1$) live in symbiosis \cite{MMM}, that is, the gain of a given group increases then the other group is also present.

Here, we consider the case where the two groups $s$ and $f$ have $0\le c_g<1$, $g=s,f$. The apparent similarity of this setup with the case previously studied \cite{MMM} is deceptive, as we shall see. The strategies are drawn according to
\bea
    P_f(a^\mu_{i,1}=a^\mu_{i,2}) &=& c_f;\\
    P_s(a^\mu_{i,1}=a^\mu_{i,2}) &=& c_s. \\      
\eea
When $c_f<c_s$, group $f$ is the fast one, and group $s$ is the slow one.
The asymmetric phase of the model for an arbitrary number of groups is exactly solvable in the limit $N\to\infty$ with replica trick \cite{MPV,CMZe00,MMM}. The solution gives
\bea
\frac{H}{N}=\frac{\phi c_f+(1-\phi)c_s+\phi(1-c_f)Q_f+(1-\phi)(1-c_s)Q_s}{(1+\phi\chi_f+(1-\phi)\chi_s)^2}
\label{eq:H(c1,c2)}
\eea
where $Q_g=\sum_{i\in {\cal N}_g}\avg{s_i(t)}^2$  ($g=f,s$); 
$\chi_f$ and $\chi_s$ are the integrated response functions \cite{CoolenBatch}. These four quantities all depend on two variables $\zeta_f$ and and $\zeta_s$ which are the solutions of two coupled non-linear equations (see appendix). Eq. \req{eq:H(c1,c2)} shows that in this case the stationary state of the asymmetric phase depends on the composition of the population, which is also true of the location of the critical point.

The respective gains of the two groups can also be computed exactly. Starting from the total losses
\bea
   \frac{\sigma^2}{N}= \frac{H}{N}+\phi(1-c_f)(1-Q_f)+(1-\phi)(1-c_s)(1-Q_s)
   \label{eq:sigma}
\eea 
and observing that if $\sigma^2_\lambda=\sum_{\mu=1}^P\avg{(A_{\rm fast}+\lambda A_{\rm slow})^2|\mu}/P$, the gains of the agents are nothing else than
\bea
   \gamma_s&=&-\avg{A_{\rm slow}A}=-\frac{1}{2}\left .\frac{\partial \sigma^2_\lambda}{\partial \lambda}\right|_{\lambda=1}\\
   \gamma_f&=&-\sigma^2-\gamma_s.
\label{eq:gains}
\eea
From the replica calculus, we find
\bea
  \frac{\gamma_s}{N_s} &=& \frac{H}{N}\chi_s-\frac{c_s+(1-c_s)Q_s}{1+\phi\chi_f+(1-\phi)\chi_s}-(1-c_s)(1-Q_s);
\label{eq:gainslow}
\eea
\bea
  \frac{\gamma_f}{N_f} &=& -\frac{1}{\phi}\frac{\sigma^2}{N}-\frac{1-\phi}{\phi}\frac{\gamma_s}{N_s}.
\label{eq:gainfast}
\eea

\begin{figure}
\centerline{\includegraphics*[width=0.37\textwidth]{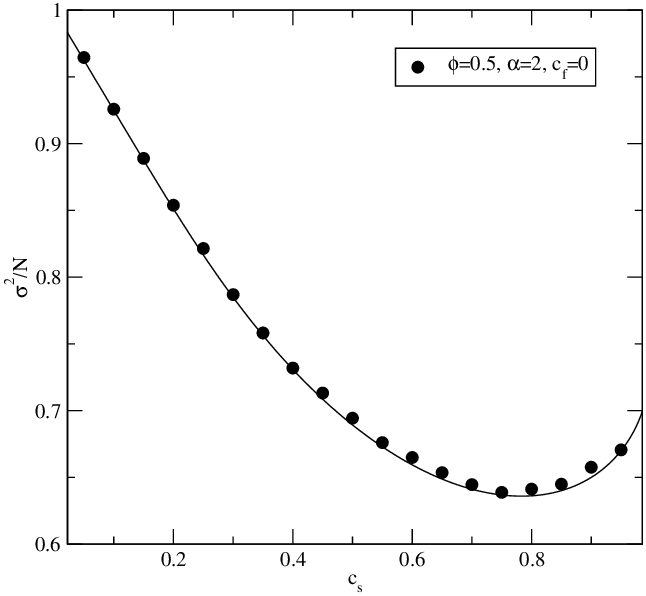}}
\caption{Fluctuations $\sigma^2/N$ versus the strategy correlation parameter $c_s$ of the slow agents ($\phi=0.5$, $c_f=0$). Simulations with $P=64$, $\alpha=2$, averages over 200 samples.}
\label{fig:article6}
\end{figure}

\begin{figure}
\includegraphics*[scale=0.7]{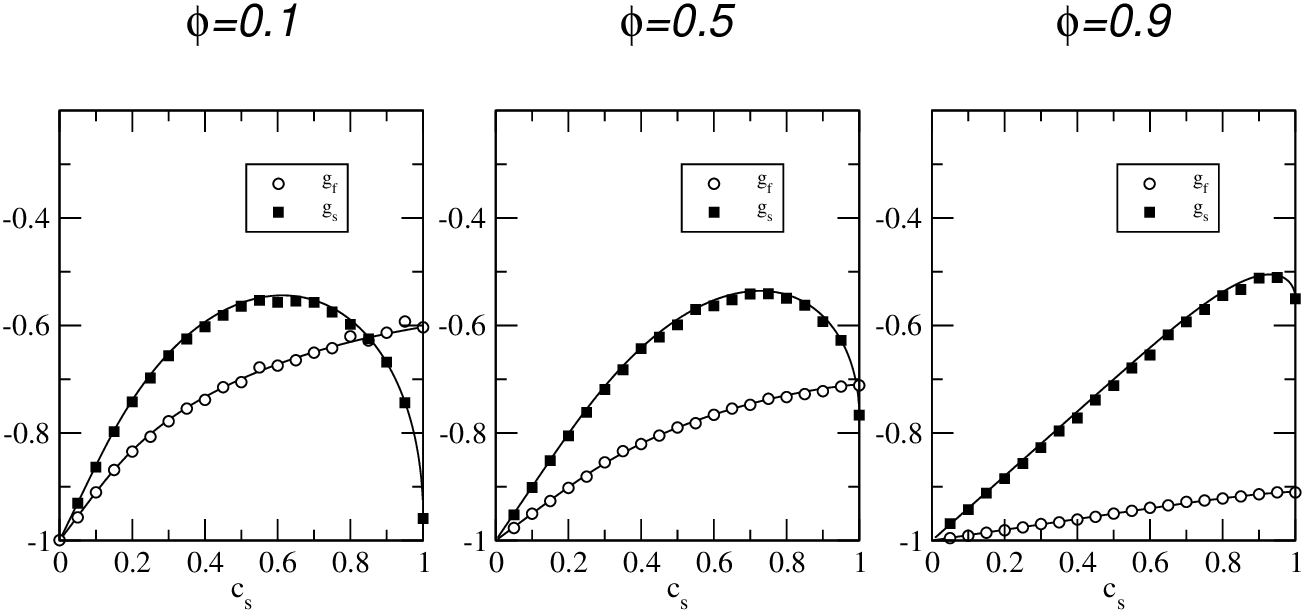}
\caption{The average gains $g_f/N_f,g_s/N_s$ for fast and slow agents respectively ($\phi=0.5, c_f=0$). Lines are theoretical predictions from Eq \req{eq:gains}. Simulations with $N=32, P=64$, averages over 200 samples}.   
\label{fig:article7}
\end{figure}

The behaviour of the game is non trivial, which is already clear in Fig \ref{fig:article6}, where the fluctuations, or average losses of the players per time-step, are plotted agains $c_s$, showing a maximum. In this kind of plot, one should be careful to stay in the asymmetric phase, because the critical point $\alpha_c$ depends on $c_s$, $c_f$, and $\phi$. Taking $\alpha>1$ solves this problem since $\alpha_c\le1$.  The minimum of $\sigma^2/N$ is surprising at first.
Figure \ref{fig:article7} gives a deeper understanding of this peculiar phenomenon by plotting the gains of the two groups. The minimum of the losses can be attributed entirely the slow agents only, who profit quite remarkably from the fast agents ($c_f=0$) unless they are very slow. This shows an competition between two effects: being slower means that one overreact less, in particular with respect to local fluctuations; inversely, being too slow makes it too difficult to react to being exploited. Interestingly, increasing $c_s$ increases monotonically the gains of group $f$.

\begin{figure}
\centerline{\includegraphics*[height=5cm]{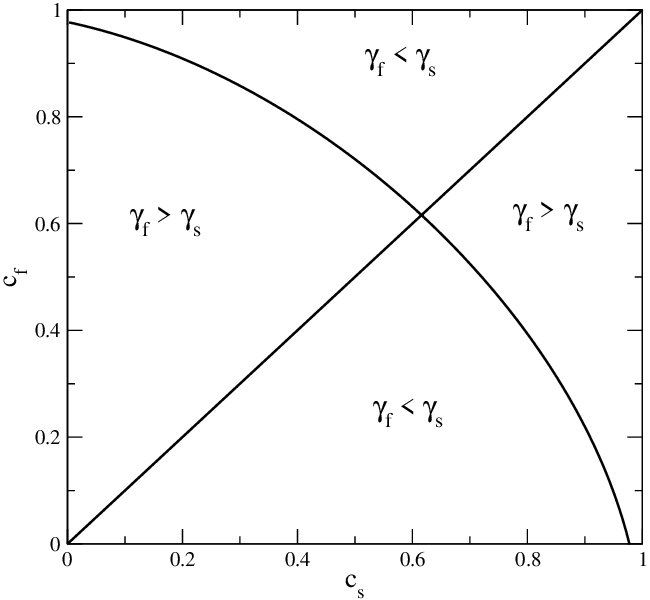}\includegraphics*[height	=5cm]{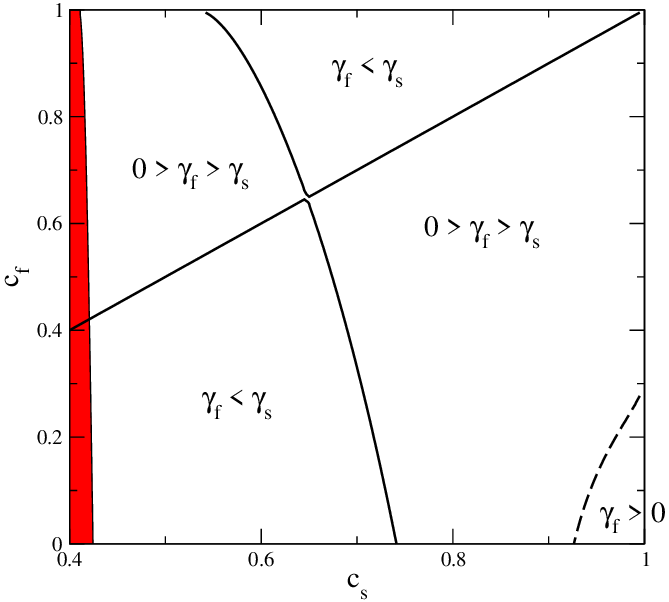}}
\caption{Regions of relative advantage for $\phi=\phi_f=0.5$ (left graph) and $\phi=0.01$ (right graph); the region filled corresponds to the symmetric phase, in which the replica calculus is not valid.}
\label{fig:gain-regions}
\end{figure}

The relative fraction of each type of agent is a crucial parameter, as illustrated by Fig \ref{fig:article7}, in particular when one group of agents has a large strategy correlation. In order to shed more light on the matter, we produced plots of the regions where one group has an advantage over the other. When $\phi=0.5$, Figure \ref{fig:gain-regions} contains four regions of interest. When $c_g<c^*\simeq 0.61$, group $g$ on average wins more than the other group $g'$ as long as $c_g>c_g'$, which means that as long as both groups have sufficiently low $c$s, the slower the better. On the other hand, when $c_g>c^(m)\simeq0.98$, group $g'$ always exploits group $g$. For intermediate values of $c_g$, the outcome depends on the precise value of both $c_g$ and $c_g'$. While changing $\phi$ leaves unaffected the diagonal boundary, it has two remarkable effects: first it changes the non-linear boundary between the $\gamma_f<\gamma_s$ and $\gamma_s<\gamma_s$ regions by roughly rotating it clockwise.  A new region also appears for $\alpha$ and $\phi$ small enough where the gain of group in relative sparseness is positive; it is however very small, and only appears when the largest group has a very large $c$. In this case, one needs to be few and react fast in order to be able to exploit very slow agents so much that one's gain is positive.

\section{Games with no public information}

The case where $P=1$, that is, when there is no public information available corresponds to the limit $\alpha=0$ of the previous section, and is particularly simple to understand analytically \cite{MC01,M01}; as a result explicit formulae for the fluctuations and gains can be obtained. Given $A(t)$, each agent $i$ receives a payoff $-a_i(t)A(t)$, which is stored in the score
\be\label{eq:Delta_i}
\Delta_i(t+1)=\Delta_i(t)-\frac{A(t)}{N}
\ee
and 
\be
P[a_i(t+1)=1]=\frac{1+\tanh(\Gamma_i\Delta_i(t))}{2}.
\ee 
$\Gamma_i$ is a reaction rate: it specifies the difference of behaviour to a change of $\Delta_i$. Because of Eq \req{eq:Delta_i} it is also timescale. 
 
Linking the above model with financial markets is straighforward if one assumes that the log-price evolves according to
\be
p(t+1)=p(t)+\frac{A}{N}.
\ee
This allows us to propose a new interpretation of this case: rewriting $\Delta_i(t)=-\sum_{t'=0}^tA(t)/N+\Delta_i(0)$ where $\Delta_i(0)$ is the initial condition of agent $i$, one sees immediately that 
\be
\Delta_i(t)=-p(t)+\Delta_i(0).
\ee
$\Delta_i(0)$ is nothing else than the asset reference price, or value, of agent $i$. Hence, this equation describes a model of $N$ investors having each a value in mind for the price, and acting accordingly.\footnote{Ref \cite{Savit} proceeded the other way around by making value investors with heterogeneous expectations on the fair value playing a delayed majority game \cite{BouchaudGiardinaCrash,dollargame} and noticing that the fluctuations look like those produced by minority players; this was shown analytically in \cite{M01}} The agents are therefore fundamentalists \cite{M01} who compare the current price with a reference price $\phi_i$, and $\Gamma_i$ tunes the price excursion from its supposed fundamental value $\Delta_i(0)$ tolerated by agent $i$.

If $\Delta_i(0)=0$, all the $\Delta_i$s are the same and can be replaced by $\Delta$. Eq \req{eq:Delta_i} becomes
\be
\Delta(t+1)=\Delta(t)-\tanh(\Gamma\Delta(t))+n(t)
\ee
where $n(t)$ is a noise term with zero average $\avg{n(t)}=0$ and
$\avg{n(t)n(t')}=\delta_{t,t'}[1-\tanh(\Gamma\Delta(t))^2]/N$; it vanishes
therefore in the $N\to\infty$ limit. It is easy to find that the fixed
point $\Delta^{(0)}=0$ is stable if $\Gamma<\Gamma^*=2$ and
$\avg{A^2}\propto N$, and unstable otherwise; in the latter case, a
period 2 dynamics emerges, with the stable points 
determined by  \cite{MC01}
\be\label{eq:Delta1}
\Delta^{(1)}=\tanh(\Gamma\Delta^{(1)})/2.
\ee
and $\avg{A^2}\propto N^2$. A Taylor expansion of Eq \req{eq:Delta1} gives $\Delta_1\simeq\pm\sqrt{\frac{3(\Gamma-2)}{\Gamma^3}}$ for $\Gamma$ close to $2$; on the other hand, $\frac{A}{2}'-\Delta_1\propto\exp(-\Gamma)$ for large $\Gamma$. A way to check numerically the value of $\Gamma^*$ is to observe the onset of the change of $\avg{\Delta^2}$ from $O(N^{-1})$ to $O(1)$ as a function of $\Gamma$.  Heterogeneous initial conditions $\Delta_i(0)\ne 0$ help to stabilize the fixed point by raising $\Gamma^*$ \cite{M01}.

\begin{figure}
\centerline{\includegraphics*[width=0.4\textwidth]{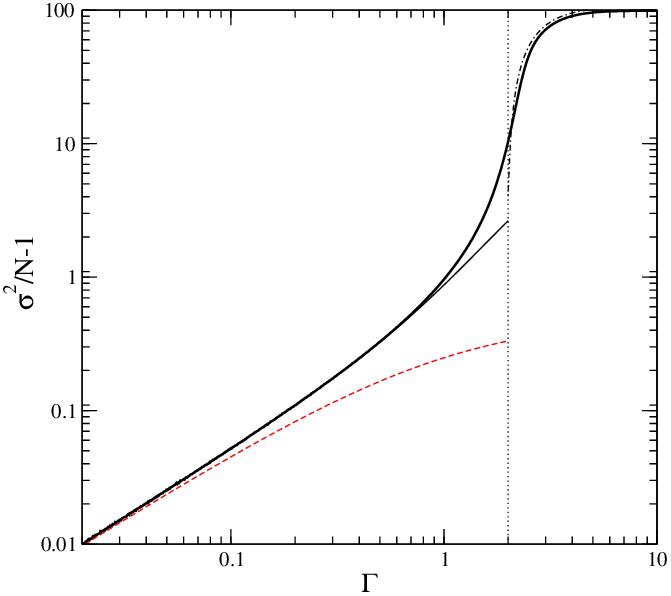}}
\caption{Fluctuations versus $\Gamma$ from numerical simulations (thick line; $N=100$, $10^6$ iterations per point), from the Fokker-Plank equation (thin line, $\Gamma<2$), the Gaussian approximation (dashed red line), and Eq \req{eq:pdeltag>2} (dash-dotted line, $\Gamma>2$)}
\label{fig:s2_vs_gamma}
\end{figure}

We shall be particularly interested in the gains of the agents. The knowledge of $P(\Delta)$ allows us to compute the average gain $\avg{\gamma}=-\avg{A^2}/N$, i.e. the fluctuations themselves
\bea
\frac{\avg{A^2}}{N}&=&\frac{1}{N}\int\dd \Delta P(\Delta)\avg{A^2|\Delta}
=\frac{1}{N}\int\dd \Delta P(\Delta)[\avg{(\delta A)^2|\Delta}+\avg{A|\Delta}^2]\label{eq:avg_gain}
\\&=&1+(N-1)\int\dd \Delta P(\Delta)\tanh(\Gamma\Delta)^2
\eea
If $\Gamma>\Gamma^*$ and $N\to\infty$, we can simply write
\be\label{fluct-largeGamma}
\frac{\avg{A^2}}{N^2}\to4\Delta_1^2
\ee
which is about $\frac{12(\Gamma-2)}{\Gamma^3}$ for $\Gamma$ close to $2$. For $\Gamma<2$, one must first keep $N$ finite and derive $P(\Delta)$.
The dynamical equation for $\Delta$ can be rewritten as a Fokker-Planck
equation, which reads

\be
\frac{\partial P(\Delta)}{\partial
  t}=\frac{\partial^2}{2\partial \Delta ^2}\cro{\frac{1-(N-1)\tanh(\Gamma\Delta)^2}{N}P(\Delta)}+\frac{\partial}{\partial \Delta}[\tanh(\Gamma\Delta)P(\Delta)]
\ee
Solving this equation in the stationary states gives
\be
P(\Delta)=\frac{1}{Z}\cro{2+N(\cosh(2\Gamma\Delta)-1))}^\frac{-(1+\Gamma)}{\Gamma}\cosh(\Gamma\Delta)^2
\ee
where $Z$ is the normalisation factor. When $\Gamma$ is small,
$P$ can be approximated by a Gaussian with zero average and $\avg{\Delta^2}=1/[2\Gamma[(N(\Gamma+1)-\Gamma]$. Therefore
\be\label{eq:s2_vs_gamma}
\frac{\avg{A^2}}{N}\simeq 1+(N-1)\frac{\Gamma}{2[N(\Gamma+1)-\Gamma]}\to 1+\frac{\Gamma}{2(\Gamma+1)}\simeq1+\frac{\Gamma}{2}~~~~~ N\to\infty,~\Gamma\ll 1
\ee 
Figure \ref{fig:s2_vs_gamma} shows that the Fokker-Planck equation provides a good description of the stationary state for $\Gamma\le 1$, whereas for larger $\Gamma$ the hypothesis of small jumps in $\Delta$ is clearly wrong; this is due to the fact that with $\Gamma>1$, the drift term makes $\Delta$ change sign on average at each time step. The Gaussian approximation bends in the wrong way as $\Gamma$ increases and should not be used for $\Gamma>0.01$. For $\Gamma>2$, $P(\Delta)$ separates into two symmetric peaks, centered roughly at $\pm \Delta_1$; it can therefore be approximated by 
\bea\label{eq:pdeltag>2}
P(\Delta)&\simeq&\frac{1}{2Z}\cro{2+N(\cosh(2\Gamma(\Delta-\Delta_1)-1))}^\frac{-(1+\Gamma)}{\Gamma}\cosh(\Gamma(\Delta-\Delta_1))^2\nonumber\\&+&\frac{1}{2Z}\cro{2+N(\cosh(2\Gamma(\Delta+\Delta_1))-1))}^\frac{-(1+\Gamma)}{\Gamma}\cosh(\Gamma(\Delta+\Delta_1))^2
\eea

\subsection{Heterogeneous learning rates} 

Let us consider $G$ groups of respectively $\phi_gN$ agents,  $g=1,\cdots,G$; all the agents belonging to group $g$ have $\Gamma_g$. Using the notation $\ovl{f(x_g)}=\sum_g\phi_gf(x_g)$ for any function $f$ of variables $x_g$, the dynamical equation for $\Delta$ reads now
\be
\Delta(t+1)=\Delta(t)-\ovl{\tanh[ \Gamma_g\Delta(t)]}+\eta(t)
\ee
where $\eta_g(t)$ are Gaussian noises with zero
average and variance
$\avg{\eta(t)\eta(t')}=\delta_{t,t'}\frac{(1-\ovl{\tanh(\Gamma_g\Delta)^2}}{N}$.
Linear stability shows that $\Delta^{(0)}$ is a stable fixed point as long as
\be\label{condstabGammagroups}
\ovl{\Gamma_g}<2.
\ee
When this point is unstable, it replaced by a period-two dynamics with stable point
\be
\Delta^{(1)}=\frac{\ovl{\tanh(\Gamma_g\Delta^{(1)})}}{2}
\ee 
Going through the same procedure as before, one finds that $P(\Delta)$ is determined by 
\be
\frac{\dd P}{P}=\frac{-2N\ovl{\tanh(\Gamma_g\Delta)}[1+\ovl{\Gamma_g(1-\tanh(\Gamma_g\Delta)^2}]-2\ovl{\Gamma_g\tanh(\Gamma_g\Delta)[1-\tanh(\Gamma_g\Delta)^2]}}{N\ovl{\tanh(\Gamma_g\Delta)}^2+1-\ovl{\tanh(\Gamma_g\Delta)^2}}
\ee
This cannot be integrated any more. The Gaussian approximation consists in keeping only the terms linear in $\Delta$ in this equation, and results in $\Delta$ being of average $0$ and variance $\avg{\Delta^2}=\frac{1}{2[N\ovl{\Gamma_g}(1+\ovl{\Gamma_g})-\ovl{\Gamma_g^2}]}$ for small $\Gamma$.

The  average gain per player of group $g$ at fixed $\Delta$ is
equal to
\be\label{gain|Delta}
\frac{\avg{\gamma_g|\Delta}}{N_g}=-\frac{\avg{A_g^2|\Delta}+\sum_{g'\ne g}{\avg{A_gA_g'|\Delta}}}{N_g}=-1-(N_g-1)\tanh(\Gamma_g\Delta)^2-\tanh(\Gamma_g\Delta)\sum_{g'}N_g'\tanh(\Gamma_g'\Delta)
\ee
Assuming that the Gaussian approximation is valid, i.e. if all the $\Gamma_g$ are small,
\be
\frac{\avg{\gamma_g}}{N_g}\simeq-1-\Gamma_g\avg{\Delta^2}\cro{-\Gamma_g+N\ovl{\Gamma_g}}\label{eq:gain_gamma}\to-1-\frac{\Gamma_g}{2[1+\ovl{\Gamma_g}]}
\ee
for infinite $N$. The gain of an agent of group $g$ can be  compared to the average gain, which yieds
\be
\avg{\gamma_g}-\ovl{\avg{\gamma_g}}\to\frac{\ovl{\Gamma_g}-\Gamma_g}{2(1+\ovl{\Gamma_g})}.
\ee 
In other words, the smaller $\Gamma_g$, the smaller the losses of that group. This is intuitive: at a given time, i.e. at a given $\Delta$, the fraction of players whose action is opposite to the sign of $\Delta$ is larger for smaller $\Gamma$.

\begin{figure}
\centerline{\includegraphics*[width=0.4\textwidth]{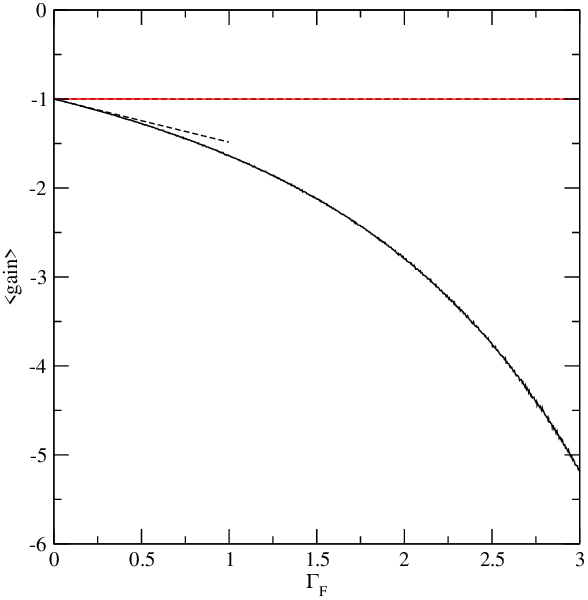}}
\caption{Average gain versus $\Gamma_f$ from numerical simulations (circles: fast agents with $\Gamma_f$, slow agents with $\Gamma_s=0.001$) and Eq \req{eq:gain_gamma} (dashed lines); average over 100000 time steps, after $10/\Gamma_s$ time steps, average over 20 samples.}
\label{fig:gamma_vs_Gamma}
\end{figure}
Figure \ref{fig:gamma_vs_Gamma} compares $\avg{\gamma}$ from numerical simulations and from Eq \req{eq:gain_gamma}. As before, the approximations made are valid for small $\Gamma$. This figure also indicates that the transition at $\Gamma^*$ is smooth.

\subsection{Heterogenous score memory $\lambda$}

Another way of having heterogeneous time scales is to differ in one's score
memory. Before studying groups, let us again characterise first the homogeneous case. For the sake of simplicity, we consider exponential moving
averages. There are two ways of implementing them:
\be\label{dynDeltaLambda}
\Delta(t+1)=\Delta(t)(1-\lambda)-\frac{A(t)}{N}
\ee 
and
\be\label{dynDeltaLambda2}
\Delta'(t+1)=\Delta'(t)(1-\lambda)-\lambda\frac{A'(t)}{N}
\ee
The length of score memory is $1/|\ln(1-\lambda)|\propto 1/\lambda$ for
small $\lambda$. One can transform Eq \req{dynDeltaLambda} into \req{dynDeltaLambda2} by setting $\Delta'=\Delta\lambda$ and $\Gamma'=\Gamma/\lambda$. The same stability analysis as above gives for a homogeneous population 
\be
\Gamma+\lambda<2
\ee
and
\be
\lambda(\Gamma'+1)<2
\ee
Increasing $\lambda$ {\em at fixed $\Gamma$ and $\Gamma'$} has therefore opposite effects for the two kinds of dynamics: a small $\lambda$ stabilises Eq \req{dynDeltaLambda} and destabilises Eq \req{dynDeltaLambda2}. Since there is a one-to-one correspondance between all the results of Eqs \req{dynDeltaLambda} and \req{dynDeltaLambda2}, we shall focus on Eq \req{dynDeltaLambda}. The period-two fixed point for $\Gamma+\lambda>2$ is now determined by $\Delta^{(1)}=\frac{1}{2-\lambda}\tanh(\Gamma\Delta^{(1)})$. Assuming that $\Delta^{(0)}=0$ is stable, the same procedure as before yields
\be
\frac{\dd P}{P}=-2N\frac{[\lambda\Delta+\tanh(\Gamma\Delta)](1+\lambda+\Gamma[1-\tanh(\Gamma\Delta)])+\Gamma\tanh(\Gamma\Delta)[1-\tanh(\Gamma\Delta)^2]}{N(\lambda\Delta+\tanh(\Gamma\Delta))^2+1-\tanh(\Gamma\Delta)^2}
\ee
\begin{figure}
\centerline{\includegraphics*[width=0.4\textwidth]{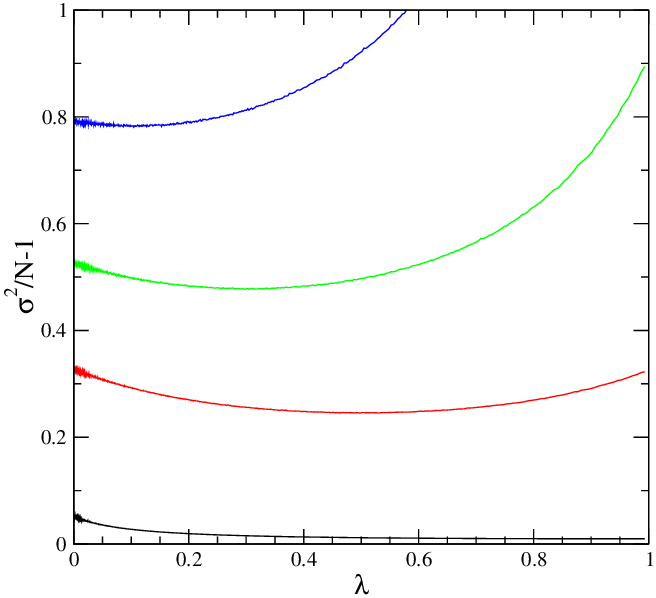}}
\caption{Fluctuations $\sigma^2/N-1$ versus $\lambda$ for $\Gamma=0.1$, $0.5$, $0.7$ and $0.9$ (bottom to top) obtained by numerical simulations. $N=100$, $10^6$ iterations per point; }
\label{fig:s2_vs_lambda}
\end{figure}
In the Gaussian approximation, one finds
\be\label{eq:D2_vs_lambda}
\avg{\Delta^2}=\frac{1}{2[(\Gamma+\lambda)(1+\lambda+\Gamma)N-\Gamma^2]}.
\ee
As consequence, for small $\Gamma$, $\lambda>0$ decreases the fluctuations. The Fokker-Planck equation holds if $\lambda+\Gamma<1$; otherwise, the change of sign of $\Delta$ at each time step (neglecting the noise) causes more fluctuations. Therefore, $\sigma^2/N$  has a minimum at $\lambda+\Gamma=1$ when $\lambda$ increases, as illustrated  by Fig \ref{fig:s2_vs_lambda}. The Gaussian approximation is very good for small $\Gamma$ and $\lambda$.

\begin{figure}
\centerline{\includegraphics*[width=0.4\textwidth]{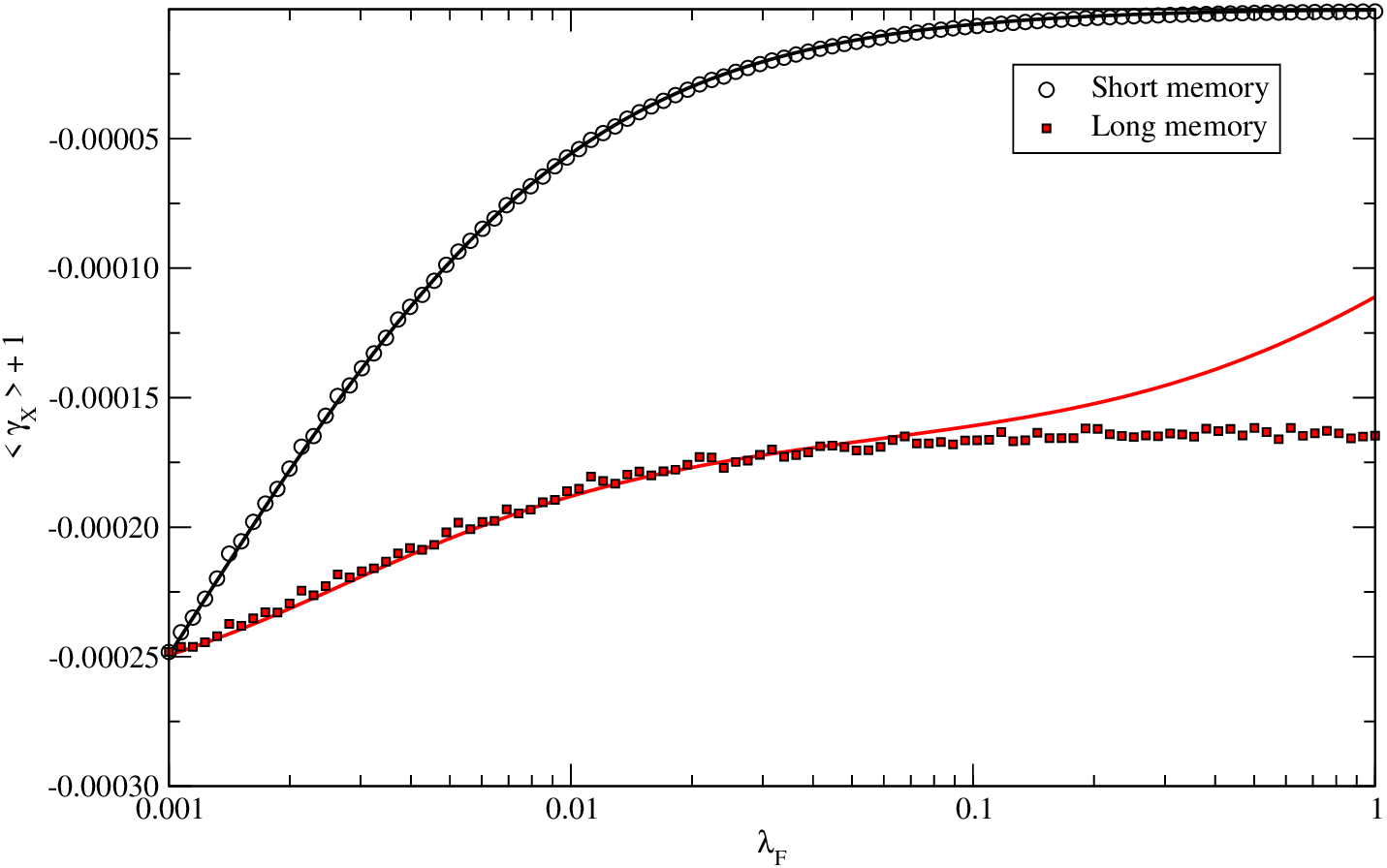}}
\caption{Average gain per time step versus $\lambda_f$ for fast players (circles) and slow players (squares). Continuous lines are from the Gaussian approximation of Eq \req{eq:gainslambda}. $\Gamma_f=\Gamma_s=0.001$, $10^{7}$ iterations per point.}
\label{fig:gain_vs_lambdaF}
\end{figure}

Generalising these results to groups of agents is more complicated than for heterogeneous $\Gamma$ because of the role of $\lambda$ in Eq. \req{dynDeltaLambda}: each group $g$ with a given $\lambda_g$ has its own $\Delta_g$
\be
\Delta_g(t+1)=\Delta_g(t)(1-\lambda_g)-\ovl{\tanh(\Gamma\Delta_{g'})}+\eta(t)=f_g(\{\Delta_{g'}\})+\eta(t)
\ee
where $\avg{\eta(t)\eta(t')}=\delta_{t,t'}(1-\ovl{\tanh(\Gamma\Delta_{g'})^2})/N$.
For infinite $N$, the linear stability conditions of the fixed point $\Delta_g=0$ for all $g$ are $\sum_{g'}|\partial f_g/\partial \Delta_{g'}|<1$, that is,
\be
|1-\lambda_g-\phi_g\Gamma_g|+\sum_{g'\ne g}\phi_{g'}\tanh(\Gamma_{g'}\Delta_{g'})<1\\
\ee
for all $g$. When $\lambda_g+\phi_g\Gamma_g<1$, i.e. both $\lambda_g$ and $\Gamma$ are small, the stability condition is $\ovl{\Gamma_{g'}}<\lambda_g+2\phi_g\Gamma_g$; for instance if $\Gamma_g=\Gamma$ for all $g$, and $\phi_g=1/G$, then the condition is $\Gamma(1-2/G)<\lambda_g$. The other case is obtained when $\Gamma_g\sim1/\phi_g$ and gives $\lambda_g+\ovl{\Gamma_{g'}}<2$, which is the same as before. Interestingly, some unstable $\Delta^{(0)}_g$ can coexist with stable ones, for instance when $\Gamma_g=\Gamma=2-\eps$, $0<\eps\ll 1$, $\phi=1/G$ and $\lambda_g=2\eps g/G$, where the dynamics of $\Delta_g$ is stable for $g<G/2$ and unstable otherwise. This is clearly a source of losses for fast-forgetting players. Going back to the average gain per player per time step, ones indeed that the gain of group $g$ is intimately related to the variance of $\Delta_g$. The multivariate Fokker-Planck equation reads
\be\label{f-p-groups-lambda}
\frac{\partial P(\Delta,t)}{\partial t}=\sum_{g}\frac{\partial}{\partial \Delta_g}\cro{\pr{[\lambda_g\Delta_g+\ovl{\tanh(\Gamma\Delta_{g'})}]P}+\frac{1}{2}\sum_{g,g'}\frac{\partial}{\partial \Delta_{g'}}(D_{g,g'}P)}
\ee
with
\be
D_{g,g'}=\frac{1-\ovl{\tanh(\Gamma_{g''}\Delta_{g''})^2}}{N}+(\lambda_g\Delta_g+\ovl{\tanh(\Gamma_{g''}\Delta_{g''})})(\lambda_{g'}\Delta_{g'}+\ovl{\tanh(\Gamma_{g''}\Delta_{g''})})
\ee
Solving a linearised version of Eq \req{f-p-groups-lambda} is done following standard procedure \cite{vK} and gives a multivariate Gaussian solution. For $G=2$, we have the resulting equations.
%
\bea
P(\Delta_f,\Delta_s)&=&\frac{1}{2\pi\sigma_f\sigma_s\sqrt{1-\rho^2}}\exp\cro{-\frac{1}{2(1-\rho^2)}\pr{\frac{\Delta_f^2}{\sigma_f^2}-\frac{2\rho\Delta_f\Delta_s}{\sigma_f\sigma_s}+\frac{\Delta_s^2}{\sigma_s^2}}}
\eea
 The expressions for $\sigma_f$, $\sigma_s$ and $\rho$ are too long to be reported here \footnote{We provide however a Mathematica file at {\tt www.maths.ox.ac.uk/~challet}}.
Reusing Eq \req{gain|Delta} the respective average gain is
\bea
\avg{\gamma_f}&\simeq&-1-\Gamma^2\pr{\avg{\Delta_f^2}(N_f-1)+N_s\avg{\Delta_f\Delta_s}}\nonumber\\
&=&-1-\Gamma_f^2\sigma_f^2(N_f-1)-\Gamma_f\Gamma_sN_s\rho\sigma_f\sigma_s]\label{eq:gainslambda}
\eea
Figure \ref{fig:gain_vs_lambdaF} plots the gains of the two groups and clearly shows that all other things beeing equal, having a shorter memory is an advantage. Indeed, when $\Gamma_f=\Gamma_s=\Gamma$ and $\phi_f=\phi_s=1/2$, as in Fig. \req{fig:gain_vs_lambdaF}, Eq \req{eq:gainslambda} leads to
\bea
\avg{\gamma_f}-\avg{\gamma_s}&\simeq&\frac{8\Gamma^2(\lambda_f - \lambda_s)(1 + \Gamma + \lambda_f +\lambda_s)[\lambda_f(1+\lambda_f+\frac{\Gamma}{2})+\lambda_s(1+\lambda_s+\frac{\Gamma}{2})+\lambda_f\lambda_s]}{ [2\lambda_f\lambda_s + \Gamma(\lambda_f + \lambda_s)][4\Gamma^2 + 
        2(2 + 2\lambda_f + \lambda_s)(2 + \lambda_f + 2\lambda_s) + 
        3\Gamma(4 + 3\lambda_f + 3\lambda_s)]}\nonumber\\
&&\times\frac{1}{2\Gamma^2 + 
        \Gamma(2 + 3\lambda_f + 3\lambda_s) + 
        2(\lambda_f + \lambda_f^2 + \lambda_s + \lambda_s^2+\lambda_f\lambda_s )}
\eea
which is of course positive when $\lambda_f>\lambda_s$. The above sections suggest that such advantage is menaced by increasing $\Gamma_f$. Interestingly, increasing $\phi_f$ increases the gains of both groups, as the fast agents suffer less from the fluctuations caused by their slow colleagues. In the unstable region, the effect is the opposite, that is, fast forgetting agents fail to smooth out sufficiently slowly large fluctuations, and suffer from larger losses than slower agents.
We performed similar numerical simulations for $P>1$ and $\Gamma=\infty$ (standard MG), and found out similar results: faster agents end up earning less in both phases (we could not find an opposite result), which can be interpreted by their tendency to switch more often between their strategies.



\section{Conclusions and remarks}

The four types of timescale heterogeneities investigated point to compatible and broad conclusions. First, agents with a smaller strategy change frequency are similar to agents with larger strategy-strategy correlation, except that their presence does not change the onset of the critical point, nor $H$.  Accordingly, the gain of fast/slow agents in both cases show similar behaviour.  Although Ref. \cite{Piai} did not compute the average gain as a function of playing frequency it did show however that agents that trade less often tend to stick more to one strategy in the asymmetric phase; since frozen agents have a higher average payoff, this suggests agent who play less often win more. We confirme this intuition by extensive numerical simulations, and checked that this conclusion does not depend on the concentration of slower agents. However, the results is the opposite in the symmetric phase where slower agents are exploited, while all the agents have an equal gain at the critical point. Overreacting, that is, having a larger reaction rate than the average population is detrimental. Reference \cite{ADMfrequency} did not study the gain of agents with heterogeneous reaction rates, therefore we refrain to generalize this conclusion to $P>1$. Finally, for $P=1$, agents with a smaller finite score memory do generally better than average, unless they have a too large reaction rate.

D.C thanks B. Hambly for technical help, and Wadham College for support.

\bibliography{timescale}

 \appendix
 \section{Replica calculus}
The calculus parallels mostly the established procedure \cite{CMZe00,MMM}: the dynamics minimizes $H$ \cite{MC01} that is akin to an energy. The stationary state of the system corresponds therefore to the ground state  of $H$.
After some algebraic manipulations it is possible to find relations that link the quantities that we have introduced The predictability is given by

\be
\frac{H_\lambda}{N}=\frac{\phi c_f+\lambda^2(1-\phi)c_s+\phi(1-c_f)Q_f+\lambda^2(1-\phi)(1-c_s)Q_s}{(1+\phi\chi_f+\lambda^2(1-\phi)\chi_s)^2}
\ee
where the integrated response functions $\chi_s$ and $\chi_f$ are defined as
\bea
&\chi&\!\!\!\!_f=\left[ \frac{\alpha-\left( 1-\phi\right) \mbox{erf}\left( \frac{\zeta_s}{\sqrt{2}}\right) }{\mbox{erf}\left( \frac{\zeta_f}{\sqrt{2}}\right) }-\phi\right]^{-1} \hspace{0.5cm},\hspace{0.5cm} \frac{\chi_s}{\chi_f}=\frac{\mbox{erf}\left( \frac{\zeta_s}{\sqrt{2}}\right) }{\mbox{erf}\left( \frac{\zeta_f}{\sqrt{2}}\right) } ;
\eea
The self-overlap $Q_g$ is equal to
\be
Q_g=1-\sqrt{\frac{2}{\pi}}\frac{e^{-\zeta_g^2/2}}{\zeta_g}-\left( 1-\frac{1}{\zeta_g^2}\right)\mbox{erf}\left( \frac{\zeta_g}{\sqrt{2}}\right)
\ee
All these quantities depend on $\zeta_s$ and $\zeta_f$, determined through
\bea
&\zeta&\!\!\!\!_f=\sqrt{\frac{\alpha\left( 1-c_f\right) }{\phi c_f+\left(1-\phi\right)c_s+\phi\left( 1-c_f\right)Q_f+\left( 1-\phi\right) \left( 1-c_s\right) Q_s }} \hspace{0.5cm},\hspace{0.5cm} \frac{\zeta_f}{\zeta_s}=\sqrt{\frac{1-c_f}{1-c_s}};\\
&\phi&\left( 1-c_f\right)\left( \sqrt{\frac{2}{\pi}}\frac{e^{-\zeta_f^2/2}}{\zeta_f}+\left( 1-\frac{1}{\zeta_f^2}\right)\mbox{erf}\left( \frac{\zeta_f}{\sqrt{2}}\right)\right)+\frac{\alpha\left( 1-c_f\right) }{\zeta_f^2}+\nonumber \\
&+&\left( 1-\phi\right) \left( 1-c_s\right)\left(  \sqrt{\frac{2}{\pi}}\frac{e^{-\zeta_s^2/2}}{\zeta_s}+\left( 1-\frac{1}{\zeta_s^2}\right)\mbox{erf}\left( \frac{\zeta_s}{\sqrt{2}}\right)\right)=1. 
\eea
Finally, the average gain of slow agents is 
\bea 
\frac{\gamma_s}{N_s}=-\frac{1}{1-\phi}\frac{1}{2}\frac{\partial}{\partial \lambda}\frac{\sigma_\lambda^2}{N}\vert_{\lambda=1}=
-\frac{1}{1-\phi}\lim_{\beta\to\infty}\left.\frac{\partial H_\lambda(\beta)}{\partial \lambda}\right\vert _{\lambda=1}-(1-c_s)(1-Q_s)
\eea
that leads to the expressions \req{eq:gainslow}.
\subsection{More than two groups}
The results above are readily generalised to $G$ groups denoted by $g=1,\dots,G$: group $g$ comprises $N_g=\phi_g N$ agents equipped with two strategies with correlation $c_{g}$:
\bea
\frac{H}{N}&=&\frac{\avg{ c+\left( 1-c\right)Q}}{\left( 1+\avg{\chi}\right ) ^2 } \\
\frac{\sigma^2}{N}&=&\frac{H}{N}+\frac{\avg{ (1-c) (1-Q)}}{N} \\
\frac{\gamma_g}{N_g}&=&\frac{H}{N}\chi_g-\frac{c_g+\left( 1-c_g\right) Q_g}{1+\avg{\chi}}-\left( 1-c_g\right) \left( 1-Q_g\right) 
\eea
where the average $\avg{.}$ is over the groups.

\bea
Q_g&=&1-\sqrt{\frac{2}{\pi}}\frac{e^{-\zeta_g^2/2}}{\zeta_g}-\left( 1-\frac{1}{\zeta_g^2}\right)\mbox{erf}\left( \frac{\zeta_g}{\sqrt{2}}\right) \\
1&=&\avg{(1-c)\pr{ \sqrt{\frac{2}{\pi}}\frac{e^{-\zeta^2/2}}{\zeta}+\pr{1-\frac{1}{\zeta^2}}\mbox{erf}\left( \frac{\zeta}{\sqrt{2}}\right)}} +\frac{\alpha( 1-c_g)}{\zeta_g^2}  \\
\frac{\zeta_g}{\zeta_{g'}}&=&\sqrt{\frac{1-c_g}{1-c_{g'}}}.
\eea 

\end{document}